Electric field lines of an arbitrarily moving charged particle

S.G. Arutunian[a,1], M.A. Aginian[a], A.V. Margaryan[a], E.G. Lazareva[a], M. Chung[b]

a) Alikhanyan National Scientific Laboratory, 0036 Yerevan, Armenia
b) Ulsan National Institute of Science and Technology, 44919 Ulsan, South Korea

**Abstract**
In this paper it is shown that the equations of electric field lines of an arbitrarily moving charged particle in the general case are reduced to homogeneous, linear differential equations with variable coefficients. For trajectories where the expression $b = \varsigma / \gamma \kappa$ is a constant ($\varsigma$ - orbit torsion, $\kappa$ - orbit curvature, $\gamma$ - Lorentz factor of a particle) these equations are reduced to homogeneous, linear differential equations with constant coefficients. This case, in particular, includes all planar trajectories. This paper presents solutions of the equations of electric field lines and corresponding illustrations both in the orbital plane and outside it for a charge moving in a flat monochromatic linearly polarized wave.

**1. Introduction**
Visualization of results has always been an important component of research activities in physics due to its ability to synthesize large amounts of data into effective graphics [1] with a visual representation of characteristic parameter relationships, which allows effectively use human creative and analytic capabilities. In [2] it is noted that looking through a numerical table takes a lot of mental effort, but information presented visually can be grasped in a few seconds, so the brain identifies patterns and makes instant subliminal comparisons. Looking at an object of study, especially from several points of view, allows one to understand it better and, most importantly, to generate new ideas related to both the study itself and the use of the object of study.

Visualization also plays a great role in the learning process [3]. Images that are created in this process are well remembered, and play an important role in the future use of acquired knowledge and the ability to make quick assessments if they are used. It is important that the absorption of information in graphic elude people in words (Bang Wong, creative director of MIT's Broad Institute, [4]).

This work is devoted to visualization of the electromagnetic field of relativistic charged particles. In electrodynamics, one of the imaging tools has been the so-called field lines. By definition, an electromagnetic field line is a line whose tangents coincide with the direction of the electromagnetic field (its electric or magnetic component) at a given point in space.

Interpretation of electromagnetic field lines has undergone a certain evolution: from Faraday, who represented these lines as force lines, because he explained the forces acting on bodies in electric and magnetic fields by the tension and pressure of force lines (according to [5]). In Maxwellian electrodynamics the field lines became only a way of visual representation of electromagnetic fields. Unsuccessful attempts to combine Maxwellian electrodynamics with force lines in the spirit of Faraday were made by J. J. Thomson (see [5]). Nevertheless, discussing ways to solve the problems of modern physics (for example, the fundamentality of the electron charge $e$, the speed of light $c$ and Planck's constant $\hbar$, renormalizations in quantum field theory, etc.) Dirac [6] turned to the idea of discrete (quantum) Faraday force lines.

The main obstacle to this was the noninvariance of the field lines, i.e., the dependence of their representation on the choice of the coordinate system. Note the work of Newcomb [7], where the problem of covariance of field lines was solved for a class of orthogonal electric and magnetic fields (see also [5]). This covariance was achieved by assigning certain drift velocities to the field lines. Covariant equations in full differentials describing the field lines were obtained. Thus, the field lines became objects which, after applying Lorentz transformations to them, retained their meaning in the new coordinate system. In [8], we managed to show how the covariant system of moving lines can be used to restore full information about the electromagnetic field tensor. In fact, this realizes the possibility to describe the electrodynamics of such orthogonal fields only by means of the field line apparatus.

Let us note one more aspect of usefulness of representation of fields of relativistic particles by means of field lines, connected with the fact that the traditional approach of Fourier decomposition of fields of radiation of relativistic charges leads to a large number of field harmonics. Thus, synchrotron radiation, contains $\gamma^3$ basic harmonics ($\gamma$ - Lorentz factor of the particle), whereas the use of Lienar-Wiechert fields and field lines allows one to obtain clear information about the spatial distribution of radiation [9]. The approach is also useful for studies of compact bunches of charged particles, when the radiation is coherent (see, for example, [10]).

In this paper we will focus on the solution of the field line equations of an arbitrarily moving point charge on a plane and consider in detail the case of electron motion in the field of a flat monochromatic linearly polarized wave. Visualization of electric field lines includes the case of relativistic motions.

Examples of charge field lines can be found, for example, in [11] (lines of electric force for a point charge embedded in a dielectric, electric field lines penetrating through the hole in conducting plane etc.), in [12] (drawings of electric field of the charge experiencing sudden stops). More complex examples of synchrotron radiation field lines in the orbital plane are given in [13]. Electromagnetic radiation fields via field lines for positive charge that has been accelerated from the rest in a short time and then proceeded with a constant velocity are considered in [14]. The problem of field lines of an arbitrarily moving charge was solved in 1983 [15] (see also [16] and [17]). In [18] the radiation field during instantaneous acceleration of a charged particle was considered in detail.

T. Shintake [19] made an interesting attempt to animate the synchrotron radiation using electric field lines in the orbital plane. The spatial structure of ribbon-like region of high electric field of synchrotron radiation is considered in [20]. It is noted that the ribbon in relativistic case is very narrow (with width in orthogonal to orbit plane proportional to $\gamma^{-1}$ and thickness proportional to $\gamma^{-3}$).

Bolotovsky and Serov [21] consider in detail the problem of the motion of a charge in the field of a plane linearly polarized electromagnetic wave. The properties of radiation of such a charged particle are also considered. Electric field of two charges instantly accelerated and moving with the constant velocity are considered in [22].

It is noted in [23, 24] that changing the electron's trajectory disturbs its eigenfield. This disturbance of the field is treated as a packet of free plane electromagnetic waves which conserves its shape in space for a long-time interval. Such localization of the field is clearly demonstrated by electric field lines.

An attempt to draw electric field lines in space was made, apparently for the first time, in [25]. The electric field lines of synchrotron radiation of a point charge with the $\gamma = 1.5$ were presented.

Let us also note the work [26], in which three cases of charged particle motion are considered: rectilinear motion, motion in the magnetic fields of wigglers/undulators, and uniform circular motion. In the case of rectilinear motion, the authors use a solution, which is a particular case of the transformations we introduced in [17] (the functions generally depending on the retarded time turn out to be constant for such motion). In the other two cases, the equations are solved analytically, but without defining the general formalism. In the presented paper we confirm again that the transformation applied in [17] solves the problem of electromagnetic field lines in the general case, including any motions on the plane (see below, including notes #2 and #3 on pages 6 and 8 for details).

## 2. Electromagnetic field lines of an arbitrarily moving charge

According to the formulas of the retarded potentials, the field of an arbitrarily moving charge $e$ at a point with a radius vector $\vec{r}$ at a time $t$ is determined by the state of motion of the charge at the retarding time moment $t'$. Let the charged particle moves along the trajectory $\vec{r}_0(t)$. Then the time moment $t'$ is defined by the retardation equation

$$c(t-t') = |\vec{r} - \vec{r}_0(t')|, \qquad (1),$$

where $c$ - the speed of light.

The electric $\vec{E}$ and magnetic $\vec{H}$ fields of an arbitrarily moving charged particle are written in the form [27].

$$\vec{E} = \frac{e\gamma^{-2}}{R^2(1-\vec{n}\vec{\beta})^3}\left\{\vec{n} - \vec{\beta} + R\gamma^2\left[\vec{n} \times \left[(\vec{n} - \vec{\beta}) \times \frac{d\vec{\beta}}{cdt'}\right]\right]\right\}, \qquad (2)$$

$$\vec{H} = \left[\vec{n} \times \vec{E}\right], \qquad (3)$$

where $R = |\vec{r} - \vec{r}_0(t')|$, $\vec{\beta}c = d\vec{r}_0(t')/dt'$, $\beta = |\vec{\beta}|$, $\gamma = 1/\sqrt{1-\beta^2}$ - Lorentz factor of a charged particle, $\vec{n} = (\vec{r} - \vec{r}_0(t'))/R$ - unit vector from the point where the charge is at the time $t'$ to the observation point $\vec{r}$. In fact, all quantities in the right-hand sides of equations (2) and (3) are taken at the time $t'$.

For an arbitrary charge trajectory, the retardation equation has no analytical solution. To avoid this difficulty, Fourier decompositions of fields (2) and (3) are used, which for ultrarelativistic motions contain a large number of harmonics (for synchrotron radiation of order $\gamma^3$).

**2.1. Electric field lines**

As noted above, the strength of the Lienard-Wiechert field at an observation point is determined by the motion characteristics of the field-creating charge at a single point of the trajectory. It is convenient, therefore, to use the notion of point light signals. A set of such signals emitted at a retarded moment of time $t'$ at the moment of observation $t$ form a sphere with radius $R = c(t-t')$ and with the center at $\vec{r}_0(t')$. The system of spheres nested into each other and not crossing each other because the velocity of the charge is always less than the speed of light serves as a convenient spatial coordinate system. An arbitrary line in space can be defined by parametrization $(R(\zeta), \vec{n}(\zeta))$ with respect to some parameter $\zeta$. Construction of lines penetrating all light spheres, for instance, can be done using as a parameter $\zeta$, the retardation time $t'$ or the radius of light spheres. Note that the electric field lines are exactly such lines since the product $\vec{E}\vec{R} = e\gamma^{-2}R^{-1}(1-\vec{\beta}\vec{n})^{-2}$ never turns to zero ($\vec{R} = R\vec{n}$).

Thus, the electric field lines at a moment of time $t$ are searched for in the following parametric form

$$\vec{L}(R) = \vec{r}_0(t - R/c) + R\vec{n}(R), \qquad (4)$$

where the $\vec{n}(R)$ (i.e., unit vector's dependence on the $R$) is not yet defined. In fact the line (4) defines a geometrical position of light signals at the moment of time $t$, which were emitted from the trajectory at time moments $t - R/c$ in the direction of the vector $\vec{n}(R)$ (we use a laboratory coordinate system $(X,Y,Z)$, and considering planar trajectories we assume that the particle moves in the plane $(X,Y)$). When changing the parameter $R$ from zero to $+\infty$ the line (4) unfolds from the point of charge location at a moment of time $t$ moving away to infinity, irrespective of whether the trajectory is closed or moves away to infinity too. Such parametrization obviously connects by the retardation equation a point in space with a radius-vector $\vec{L}(R)$ with the position of the charge at a moment of time $t' = t - R/c$.

The derivative of the line $\vec{L}(R)$ by $R$ defines a tangent to this line and is written in the form

$$\frac{d\vec{L}}{dR} = -\vec{\beta} + \vec{n} + R\frac{d\vec{n}}{dR}. \tag{5}$$

If the calculated tangent to the line $\vec{L}(R)$ is proportional to the electric field at the point $\vec{L}(R)$, then this line is the electric field line.

This is so, for example, if the right-hand side of equation (5) coincides with the curly bracket in the right-hand side of equation (2). From this condition we have[1]:

$$\frac{d\vec{n}}{dR} = -\gamma^2 \left[ \vec{n} \times \left[ (\vec{n} - \vec{\beta}) \times \frac{d\vec{\beta}}{dR} \right] \right]. \tag{6}$$

Let us write down the equation (6) on components in the coordinate system accompanying to the trajectory of motion at the moment of time $t - R/c$ with the following unit vectors: $\vec{e}_2$ - vector in the direction of particle's velocity $\vec{\beta}$, vector $\vec{e}_1$ in opposite direction to trajectory normal (when unfolding it in the direction of particle's motion), $\vec{e}_3 = [\vec{e}_1 \times \vec{e}_2]$.

We write the vector $\vec{n}$ in the following form

$$\vec{n} = n_1 \vec{e}_1 + n_2 \vec{e}_2 + n_3 \vec{e}_3. \tag{7}$$

Let us use the Frenet-Serret relations defining differentiation of unit vectors of the accompanying coordinate system along the trajectory arc length $ds$ [28] (when differentiating we assume that the trajectory unfolds in the direction of particle's motion):

$$\frac{d\vec{e}_1}{ds} = \kappa \vec{e}_2 - \varsigma \vec{e}_3, \tag{8}$$

$$\frac{d\vec{e}_2}{ds} = -\kappa \vec{e}_1, \tag{9}$$

$$\frac{d\vec{e}_3}{ds} = \varsigma \vec{e}_1, \tag{10}$$

where $\kappa$ is the curvature and $\varsigma$ is torsion of the trajectory.

---

[1] A component notation of this vector equation was given in [15].

Substituting these relations into the left-hand side of equation (6), we obtain the following relation (given that $ds = -\beta dR$):

$$\frac{d\vec{n}}{dR} = \vec{e}_1 \left\{ \frac{dn_1}{dR} + \beta \kappa n_2 - \beta \varsigma n_3 \right\} + \vec{e}_2 \left\{ \frac{dn_2}{dR} - \beta \kappa n_1 \right\} + \vec{e}_3 \left\{ \frac{dn_3}{dR} + \beta \varsigma n_1 \right\}. \tag{11}$$

Given also that:

$$\frac{d\vec{\beta}}{dR} = \frac{d\beta}{dR} \vec{e}_2 + \beta^2 \kappa \vec{e}_1 \tag{12}$$

we can describe the right-hand side of equation (6) as

$$-\gamma^2 \left[ \vec{n} \times \left[ (\vec{n} - \vec{\beta}) \times \frac{d\vec{\beta}}{dR} \right] \right] = \vec{e}_1 \left\{ -\beta^2 \gamma^2 \kappa (1 - n_1^2) - \beta^3 \gamma^2 \kappa n_2 - \frac{d\beta}{dR} \gamma^2 n_1 n_2 \right\} +$$

$$+ \vec{e}_2 \left\{ -\beta^2 \gamma^2 \kappa n_1 (n_2 - \beta) + \frac{d\beta}{dR} \gamma^2 (1 - n_2^2) \right\} + \tag{13}$$

$$+ \vec{e}_3 \left\{ -\beta^2 \gamma^2 \kappa n_1 n_3 - \frac{d\beta}{dR} \gamma^2 n_2 n_3 \right\}$$

The final result is the equations [15]:

$$\frac{dn_1}{dR} = -\beta \gamma^2 \kappa n_2 + \beta \varsigma n_3 + \beta^2 \gamma^2 \kappa (1 - n_1^2) - \frac{d\beta}{dR} \gamma^2 n_1 n_2, \tag{14}$$

$$\frac{dn_2}{dR} = \beta \gamma^2 \kappa n_1 - \beta^2 \gamma^2 \kappa n_1 n_2 + \frac{d\beta}{dR} \gamma^2 (1 - n_2^2), \tag{15}$$

$$\frac{dn_3}{dR} = -\beta \varsigma n_1 - \beta^2 \gamma^2 \kappa n_1 n_3 - \frac{d\beta}{dR} \gamma^2 n_2 n_3. \tag{16}$$

To solve equations (14)-(16), we replace the unknown functions $n_1(R), n_2(R), n_3(R)$ with functions $v_1(R), v_2(R), v_3(R)$ of the form

$$n_1 = \frac{v_1}{\gamma(1 + \beta v_2)}, \tag{17}$$

$$n_2 = \frac{v_2 + \beta}{1 + \beta v_2}, \tag{18}$$

$$n_3 = \frac{v_3}{\gamma(1+\beta v_2)}. \tag{19}$$

Note that transformations (17) - (19) can be written in vector form[2] [17]

$$\vec{n} = \frac{\vec{\beta}(1+(\vec{\beta}\vec{v}(1-\gamma^{-1})/\beta^2))+\vec{v}\gamma^{-1}}{1+\vec{\beta}\vec{v}}. \tag{20}$$

It is interesting to note that the replacement (20) is a Lorentz transformation of the speed $\vec{n}c$ of the light signal from the laboratory reference frame to the speed of the light signal $\vec{v}c$ in the accompanying inertial reference frame associated with the trajectory at the retarded point of the trajectory $\vec{r}_0(t-R/c)$.

Transforms (17)-(19) should be substituted into equations (14)-(16) to obtain equations for the $v_1(R), v_2(R), v_3(R)$ functions.

Let us start with equation (15) for $n_2$. We have for the left-hand side of this formula (hereafter the prime denotes the derivative by $R$):

$$n_2' = \left(\frac{v_2+\beta}{1+\beta v_2}\right)' = (1+\beta v_2)^{-2}\{(\beta'+v_2')(1+\beta v_2)-(\beta+v_2)(\beta' v_2+\beta v_2')\} =$$
$$(1+\beta v_2)^{-2}\{v_2'(1-\beta^2)+\beta'(1-v_2^2)\} \tag{21}$$

Further, by substituting formulas (17) and (18) into the right-hand side of equation (15), we obtain

$$\beta\gamma^2\kappa\frac{v_1}{\gamma(1+\beta v_2)}\left(1-\frac{\beta(v_2+\beta)}{1+\beta v_2}\right)+\beta'\gamma^2\left(1-\frac{(v_2+\beta)^2}{(1+\beta v_2)^2}\right) =$$
$$(1+\beta v_2)^{-2}\{\beta\gamma\kappa v_1(1-\beta^2)+\beta'(1-v_2^2)\} \tag{22}$$

By equating (21) and (22), we have

$$(1+\beta v_2)^{-2}\{v_2'(1-\beta^2)+\beta'(1-v_2^2)\} = (1+\beta v_2)^{-2}\{\beta\gamma\kappa v_1(1-\beta^2)+\beta'(1-v_2^2)\}$$

Or we finally get the equation for $v_2$:

$$\frac{dv_2}{dR} = \beta\gamma\kappa v_1 \tag{23}$$

---

[2] In [26] conversions (17) and (18) were used to solve equations (14), (15) in the case of rectilinear motion of particles (the problem was considered in plane $(x,y)$). In this case functions $(v_1, v_2)$ turn to constants (see also the note #3 on page 8).

For $n_1$ from equation (17) we have:

$$n_1' = \left(\frac{v_1}{\gamma(1+\beta v_2)}\right)' = \frac{v_1'}{\gamma(1+\beta v_2)} - \frac{v_1(\gamma'+(\beta\gamma)'v_2)}{\gamma^2(1+\beta v_2)^2} - \frac{\beta\gamma v_1 v_2'}{\gamma^2(1+\beta v_2)^2}$$

Taking into account (23) as well as the relation:

$$\gamma' + (\beta\gamma)'v_2 = \beta'\gamma^3(\beta+v_2)$$

we rewrite equation (14) as

$$\frac{v_1'}{\gamma(1+\beta v_2)} = -\frac{\beta\gamma^2\kappa(\beta+v_2)}{1+\beta v_2} + \frac{\beta\varsigma v_3}{\gamma(1+\beta v_2)} + \beta^2\gamma^2\kappa$$

Or finally for $v_1$ have:

$$\frac{dv_1}{dR} = -\beta\gamma\kappa v_2 + \beta\varsigma v_3, \tag{24}$$

Performing a similar calculation for equation (16), we obtain the equation for $v_3$

$$\frac{dv_3}{dR} = -\beta\varsigma v_1 \tag{25}$$

Equations (23)-(25) can be rewritten in vector form [3]:

$$\frac{d\vec{v}}{dR} = \frac{\gamma-1}{\beta^2}\left[\vec{v}\times\left[\vec{\beta}\times\frac{d\vec{\beta}}{dR}\right]\right]. \tag{26}$$

Formally, equation (26) describes the rotation of a vector $\vec{v}$ with instantaneous angular velocity

$$\vec{\Omega} = (\gamma-1)\left[\vec{\beta}\times d\vec{\beta}/dR\right]/\beta^2. \tag{27}$$

Introducing a new differentiation variable

$$d\varphi = -\beta\gamma\kappa\, dR \tag{28}$$

and having marked

---

[3] It is easy to see that for rectilinear trajectories the vectors $\vec{\beta}$ and $d\vec{\beta}/dR$ are parallel, so that the right-hand side of (26) is zero. In this case the functions $v_1, v_2, v_3$ become constant, which corresponds to the results of [26].

$$b = \varsigma / \gamma \kappa \tag{29}$$

equations (23)-(25) are rewritten as a system of linear homogeneous equations:

$$\frac{dv_1}{d\varphi} = v_2 - bv_3, \tag{30}$$

$$\frac{dv_2}{d\varphi} = -v_1, \tag{31}$$

$$\frac{dv_3}{d\varphi} = bv_1. \tag{32}$$

It is easy to see that equations (30) - (32) have an obvious integral $v_1^2 + v_2^2 + v_3^2 = const$ (the constant should be put equal to unity to ensure $|\vec{n}| = 1$). Equations (30) - (32) for non-constant coefficients are solved, for example, by the method of successive approximations [29].
In the case of $b = const$ equations (30)-(32) reduce to a system of linear homogeneous equations with constant coefficients, which are easily solved.
Let us rewrite equations (30) - (32) in matrix form

$$\frac{d}{d\varphi}\begin{pmatrix} v_1 \\ v_2 \\ v_3 \end{pmatrix} = \begin{pmatrix} 0 & 1 & -b \\ -1 & 0 & 0 \\ b & 0 & 0 \end{pmatrix} \begin{pmatrix} v_1 \\ v_2 \\ v_3 \end{pmatrix} \tag{34}$$

Let us find partial solutions to this system in the form of

$$\begin{pmatrix} v_1 \\ v_2 \\ v_3 \end{pmatrix} = \exp(\lambda\varphi) \begin{pmatrix} c_1 \\ c_2 \\ c_3 \end{pmatrix}, \tag{35}$$

where $\lambda$ is the root of the characteristic equation of the system (34):

$$\lambda(\lambda^2 + 1 + b^2) = 0. \tag{36}$$

For each root of the characteristic equation, the corresponding eigenvector $\begin{pmatrix} c_1 \\ c_2 \\ c_3 \end{pmatrix}$ should be also found.

From (36) we have:

$$\lambda_1 = 0, \quad \lambda_{2,3} = \pm i\sqrt{1 + b^2}. \tag{37}$$

Finding the corresponding eigenvectors, we obtain a general solution of equations (34):

$$v_1 = H\sqrt{1+b^2}\sin(\sqrt{1+b^2}(\varphi-\varphi_0)), \tag{38}$$

$$v_2 = H\cos(\sqrt{1+b^2}(\varphi-\varphi_0)) \pm b\sqrt{1/(1+b^2)-H^2}, \tag{39}$$

$$v_3 = -bH\cos(\sqrt{1+b^2}(\varphi-\varphi_0)) \pm \sqrt{1/(1+b^2)-H^2}. \tag{40}$$

The solution depends on two integration constants $\varphi_0$ and $H$ ($\varphi_0$ varies within $(0, 2\pi)$, $H$ varies within $(0, 1/\sqrt{1+b^2})$). For relations (38)-(40), the condition $v_1^2+v_2^2+v_3^2=1$ is fulfilled. From equations (39)-(40) the integral of motion also follows

$$bv_2 + v_3 = \pm\sqrt{1+b^2}\sqrt{1-H^2(1+b^2)}. \tag{41}$$

Equations (41) in the plane $(v_2, v_3)$ define a set of straight lines parallel to the line $bv_2 + v_3 = 0$, which corresponds to the condition $H = 1/\sqrt{1+b^2}$. It becomes clear that the lines of electric field correspond to the motion of the point along the circles in the phase space $(v_1, v_2, v_3)$, which are defined by the intersection of the set of planes orthogonal to the plane $(v_2, v_3)$ and passing through the lines corresponding to the condition (41) with the unit sphere with the center at the origin of coordinates. The condition $H = 1/\sqrt{1+b^2}$ corresponds to a great circle of the sphere (i.e., the secant plane passes through the center of the sphere), with $H = 0$ the above planes touching the unit sphere at the opposite ends of the diameter.

An interesting special case of equations (38)-(40) is the condition $b = 0$, which defines a class of arbitrary planar trajectories of charged particles (torsion of such trajectories $\varsigma = 0$). In this case

$$v_3 = \pm\sqrt{1-H^2} = const. \tag{42}$$

The class of lines of force lying on the plane of motion of a particle is defined by the condition $H = 1$, and equations (38)-(39) for such lines of force are written in the form

$$v_1 = \sin(\varphi-\varphi_0), \tag{43}$$

$$v_2 = \cos(\varphi-\varphi_0). \tag{44}$$

### 3. Electric field lines of a charge moving in a linearly polarized plane wave

As an example, consider the electric field lines of an electron moving in the field of a plane linearly polarized wave.

### 3.1 Trajectory of a particle

Trajectory of the electron in a plane monochromatic linearly polarized wave is solved in a general form (see e.g. [27]). In a reference frame in which the electron on the average is at rest the parametric representation of motion is written in the form

$$x = -\frac{e^2 E^2 c}{8\Gamma^2 \omega^3} \sin 2\eta, \tag{45}$$

$$y = -\frac{eEc}{\Gamma \omega^2} \cos \eta, \tag{46}$$

$$z = 0, \tag{47}$$

$$t = \frac{\eta}{\omega} - \frac{e^2 E^2}{8\Gamma^2 \omega^3} \sin 2\eta, \tag{48}$$

where $\omega$ is the frequency of the wave, $\eta$ the parameter along which the trajectory unfolds, and $\Gamma^2 = m^2 c^2 + (e^2 E^2 / 2\omega^2)$ ($m$ is electron mass, $e$ is charge of electron). Electric field $\vec{E}$ in the wave is chosen in the direction of the axis $Y$: $E_y = E \cos(\omega(t - x/c))$, the wave propagates in the direction of the axis $X$.

To have a better representation of the trajectory for various parameters of the wave field, it is convenient to rewrite the trajectory of motion using dimensionless parameters: $\xi_x = x / \lambdabar$, $\xi_y = y / \lambdabar$, $\tau = \omega t$, where $\lambdabar = c / \omega$. Note that equations (45)-(48) describe the trajectory corresponding to $\xi_x = 0$, $-\xi_y = \max$ at time $\tau = 0$.

Let us also introduce a dimensionless parameter characterizing the wave which is proportional to the ratio of electric field to its frequency:

$$\alpha = \frac{eE}{\sqrt{2} mc\omega} = \frac{eE\lambdabar}{\sqrt{2} mc^2} = \frac{1}{\sqrt{2}} \left(\frac{E}{E_0}\right) \frac{\lambdabar}{r_0}, \tag{49}$$

where $E_0 = m^2 c^4 / e^3 = (mc^2 / e) / r_0 = 1.813393 \times 10^{18}$ V/cm ($r_0$ is classical radius of the electron). The physical meaning of formula (49) is the ratio of the work done by the electric field of the wave at distances on the order of a wavelength to the rest energy of the electron.

In these units formulas (45)-(48) are rewritten as

$$\xi_x = -\frac{\alpha^2}{4(1+\alpha^2)} \sin 2\eta, \tag{50}$$

$$\xi_y = -\frac{\sqrt{2}\alpha}{\sqrt{1+\alpha^2}} \cos \eta, \tag{51}$$

$$\tau = \eta - \frac{\alpha^2}{4(1+\alpha^2)} \sin 2\eta. \tag{52}$$

In order to build electric field lines, we will also need expressions for the electron velocity components $\beta_x, \beta_y$ of the absolute velocity $\beta$, as well as the Lorentz factor of the electron $\gamma = 1/\sqrt{1-\beta^2}$:

$$\beta_x = \frac{d\xi_x}{d\tau} = -\frac{B \cos 2\eta}{1 - B \cos 2\eta}, \tag{53}$$

$$\beta_y = \frac{d\xi_y}{d\tau} = -\frac{A\sin\eta}{1-B\cos 2\eta}, \tag{54}$$

$$\beta = \sqrt{\beta_x^2 + \beta_y^2}, \tag{55}$$

$$\gamma = 1/\sqrt{1-\beta^2}, \tag{56}$$

where $A = \sqrt{2}\alpha/\sqrt{1+\alpha^2}$, $B = \alpha^2/2(1+\alpha^2)$.

The curvature of the electron trajectory in dimensionless units $\chi = \lambdabar\kappa$ is calculated by the formula [28]

$$\chi = \frac{1}{\beta^3}\left(\beta_x \frac{d^2\xi_y}{d\tau^2} - \beta_y \frac{d^2\xi_x}{d\tau^2}\right), \tag{57}$$

where

$$\frac{d^2\xi_x}{d\tau^2} = \frac{2B\sin 2\eta}{(1-B\cos 2\eta)^3},$$

$$\frac{d^2\xi_y}{d\tau^2} = \frac{A(1-B-2B\sin^2\eta)\cos\eta}{(1-B\cos 2\eta)^3}.$$

Note that all the values given in formulae (50)-(57) depend on a single parameter $\alpha$.

Figs. 1-3 show several trajectories for different values of the parameter $\alpha$ (0.1; 1; 10).

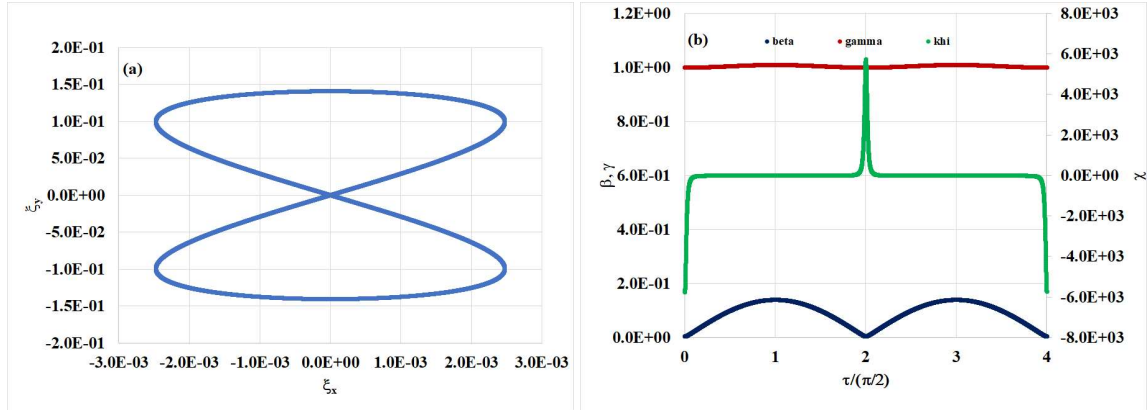

Fig. 1. (a) Trajectory of the charge in the field of a plane monochromatic linearly polarized wave in the case of $\alpha = 0.1$. The horizontal axis is directed along the wave propagation. The scales of horizontal and vertical axes are different. (b) Dependence of the velocity absolute value (blue line), the Lorentz factor $\gamma$ (red line) and the curvature $\chi$ (green line) of the charge trajectory from a dimensionless time. At the moment $\tau = 0$ the particle passes the lowest point of the trajectory ($\xi_x = 0$, $\xi_y = -0.14$). At this point the absolute value of trajectory curvature is maximal ($\chi = -5.741919E+03$), and $\beta = 4.975124E-03$, $\gamma = 1.000012E+00$ are minimal. At the moment $\tau = \pi/2$ the particle passes through the point of intersection

of the "figure of eight", at that $\chi = 0$, and the values $\beta$, $\gamma$ reach their maximum values: $\beta = 1.401129E-01$, $\gamma = 1.009963E+00$.

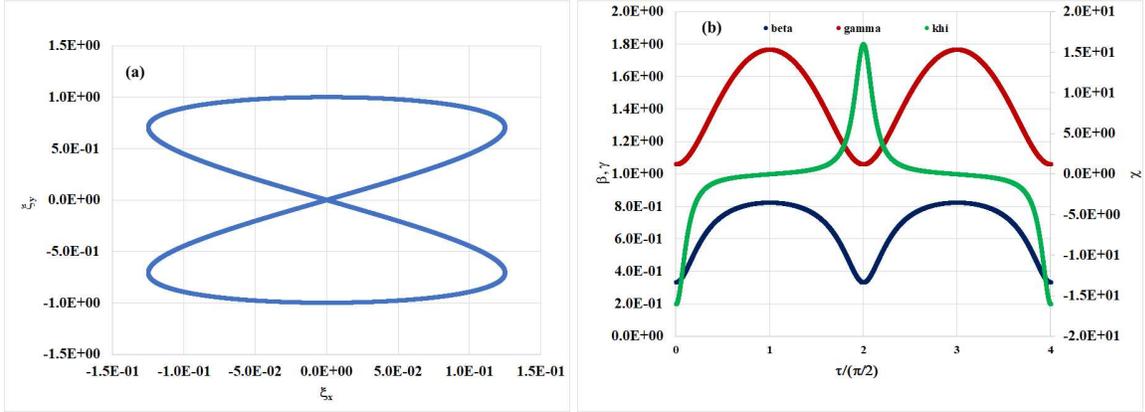

Fig. 2. Same as Fig. 1 for $\alpha =1$. For $\tau = 0$ the values $\chi = -1.600000E+01$, $\beta = 3.333333E-01$, $\gamma = 1.060660E+00$. For $\tau = \pi/2$ the values $\chi = 0$, $\beta = 8.246211E-01$, $\gamma = 1.767767E+00$.

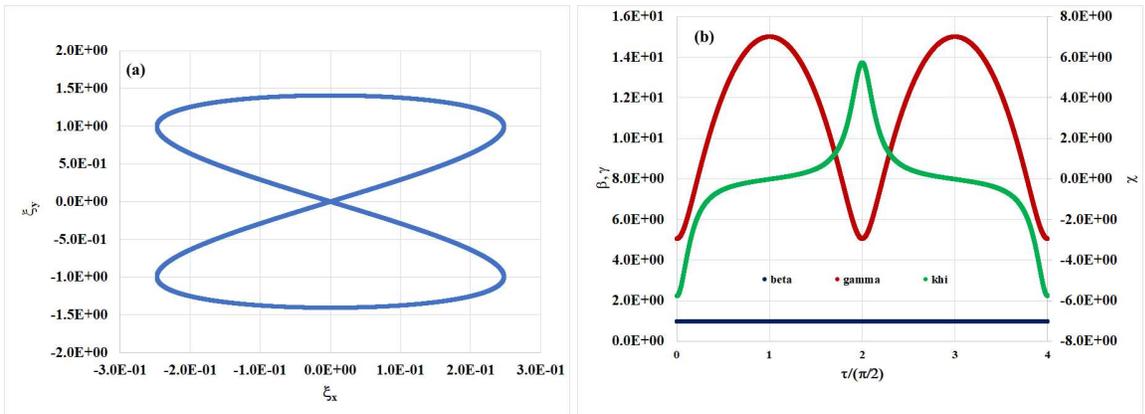

Fig. 3. Same as Fig. 1 for $\alpha =10$. For $\tau = 0$ the values $\chi = -5.741919E+00$, $\beta = 9.803922E-01$, $\gamma = 5.074690E+00$. For $\tau = \pi/2$ the values $\chi = 0$, $\beta = 9.977827E-01$, $\gamma = 1.502506E+01$.

Note that in the given units when $\alpha \to \infty$ the form of the trajectory tends to a curve $16\xi_x^2 + \xi_y^4 - 2\xi_y^2 = 0$ (according to (50)-(51)), thus starting from values $\alpha \geq 1$ the particle motion becomes relativistic (for $\alpha = 1$ parameter $\gamma$ reaches maximum value 1.768 for $\tau = \pi/2$ and minimum value 1.061 for $\tau = 0$. For all trajectories the maximal value of the absolute value of curvature radius is reached at points $\tau = 0$, $\pm\pi$, $\pm 2\pi$, ...; at points $\tau = \pm\pi/2$, $\pm 3\pi/2$, ... the trajectory changes the sign of curvature and here the parameter $\chi = 0$.

An important role in construction of electric field lines is the parameter of dynamic phase in trigonometric functions in formulas (38)-(40) (for plane trajectories in formulas (43)-(44)). This parameter characterizes the line turns around a particle and, as pointed in [17], defines the field form in space. For synchrotron radiation, where the trajectory curvature is constant, this parameter grows continuously with increasing radius of light spheres (increasing retardation depth). In the case of plane wave motion (when the motion is described by a closed figure of eight) the phase $\varphi$ is a sign-variable quantity with an amplitude dependent on the field intensity (parameter $\alpha$).

As we mentioned above, the field at the time of observation is created by the retarded positions of the particle on the trajectory. Usually, we choose $\tau = 0$ as the observation time moment, so that the retarded times are negative. Fig. 4 shows the dependence of the parameter $\varphi$ according to formula (28) on the retardation time.

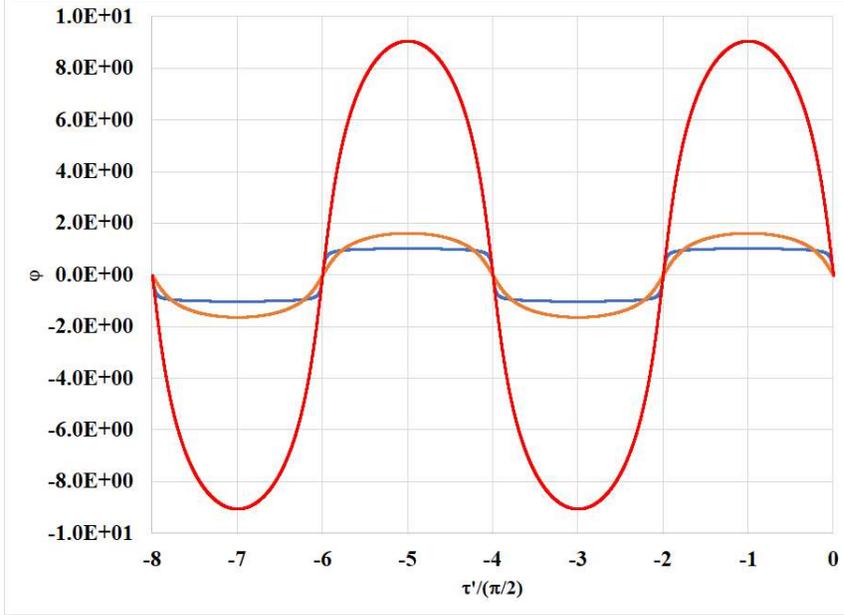

Fig. 4. Dependence of the parameter $\varphi$ on the retardation time $\tau'$ for three values of the parameter $\alpha$: 0.1 (blue line), 1 (orange line) and 10 (red line). Momentum of observation time $\tau = 0$.

From Fig. 4 it is seen that for $\alpha = 0.1$ the phase oscillates in range (-1.0, +1.0), for $\alpha = 1$ respectively within- (-1.6, 1.6), for $\alpha = 10$ within- (-9.0, 9.0).
It is interesting to consider how the total radiation intensity of a particle depends on the position of the particle on the trajectory. For the instantaneous radiation intensity of the particle, we have [27]

$$I = \frac{2}{3}e^2 c \beta^4 \gamma^4 \kappa^2 = \frac{2e^2 c}{3\lambda^2}\beta^4 \gamma^4 \chi^2. \tag{58}$$

Substituting the time dependence $\beta, \gamma, \chi$ into this formula we obtain a graphic of the dependence we are looking for (Fig. 5). Taking into account that the picture of the field we are interested in is created at retarding moments of time in the same way as in Fig. 4, the time axis is given in units of the retardation depth.

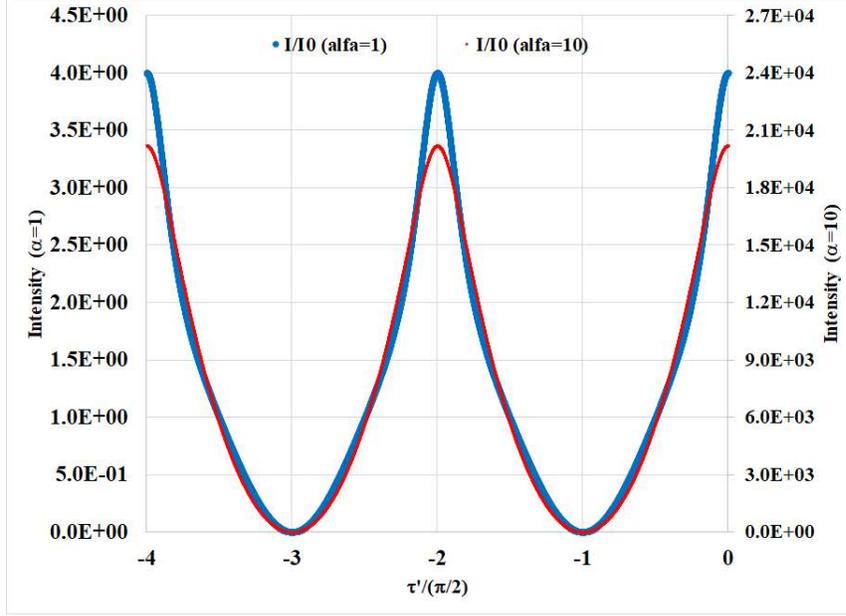

Fig. 5. Dependence of the instantaneous particle emission on the retarded time $\tau'$ at different field amplitudes $\alpha$ ($\alpha = 1$ blue line, $\alpha = 10$ red line) in units of $I_0 = 2e^2 c / 3 \lambdabar^2$. Momentum of observation time $\tau = 0$.

The maximum intensity of the radiation is observed at moments of time $\tau' = 0, \pm \pi, \pm 2\pi, \ldots$ (moments $\tau' = 0, -\pi, -2\pi$ are seen on Fig. 5) when the electron is at the tops of the figure eight. At these moments the velocity of the particle is directed against the axis $X$, and this direction is the same irrespective of the sign of the charge. For relativistic particles the radiation is directed along the velocity of the particle and therefore in the opposite direction of the wave propagation (see Fig. 11).

It is interesting to estimate the maximum value of the parameter $\alpha$ for electron moving in existing extreme laser waves. Extreme laser field intensities are successfully achieved by the Center for Relativistic Laser Science (CoReLS) [30]. In April 2021, they have achieved the record-breaking milestone of $10^{23}$ W/cm$^2$ by tightly focusing the multi-PW laser beam. Several special techniques have been employed to achieve this feat. The power intensity was maximized by using a focusing optics called an off-axis parabolic mirror, which was used to focus a 28 cm laser beam down to a spot only 1.1 micrometers wide. The CoReLS 4-PW laser is a femtosecond, ultrahigh power Ti:sapphire laser, based on the chirped pulse amplification (CPA) technique [31]. It is noted that such laser intensity enables to explore novel physical phenomena occurring under extreme physical conditions [32]. In particular, relativistic laser-plasma interactions with the possibility of laser-driven charged particle acceleration are discussed [33]. Here as an assistive task a single particle motion in a plane electromagnetic wave is considered in details. The difference between the particle motion in an infinite plane wave and in a finite laser pulse is pointed out and the formula for energy gain of the particle in laser wakefield acceleration is given. Works on laser acceleration are now on the research front in physics (see for example [34, 35]).

The achieved laser field intensity $10^{23}$ W/cm$^2$ corresponds to an electric field wave amplitude of $E \sim$ 8.7E+12 V/cm which for wavelength 800 nm gives an $\alpha = 153$ ($\gamma \sim 229$), i.e. in such a field the motion of the electron is essentially relativistic.

To calculate the amplitude of the wave field we have used the relation connecting average intensity $I$ with the maximum field strength of the electric field $E$:

$$I[W/cm^2] = \frac{1}{8\pi}\left(\frac{mc^2[J] \cdot c[cm/s]}{(r_0[cm])^3}\right)\left(\frac{E[V/cm]}{E_0[V/cm]}\right)^2 = 1.327 \cdot 10^{-3} \cdot \left(E[V/cm]\right)^2. \qquad (59)$$

It is interesting to note that a particle can also become relativistic in low frequency waves for correspondingly smaller field amplitudes.

### 3.2 Field lines in orbit plane

Let us construct electric field figures for the motion of an electron in the field of a plane wave at several values of the parameter $\alpha$. The motion and time are described in dimensionless units of formulae (50)-(52). The lines are unwound by changing the parameter $\eta$ from zero (the time moment of observation $\tau$ = 0) in the negative direction. The limit value of the parameter $\eta$ is chosen so that the retardation depth $\tau$ reaches a value of -5 in all cases. The scale of all figures is chosen to be the same. Since the sweep of lines follows strongly nonlinear formulas (especially for relativistic motions) the sweep step $d\eta$ is chosen not to be constant, but to depend on the real shift of the line point in space, so that in the screen plane this shift corresponds to approximately one pixel:

$$d\eta = \begin{cases} k_1 \cdot d\eta, & \text{if } dist \cdot scl > 1.5 \\ k_2 \cdot d\eta, & \text{if } dist \cdot scl < 0.75 \end{cases}, \qquad (60)$$

where $k_1 = 0.7$, $k_2 = 1.2$ are empirically chosen coefficients, $dist$ - maximal value of projections of line increments in real 3-dimensional space along axes $(X, Y, Z)$, $scl$ - scaling factor of line representation on the screen plane.

To begin with, we draw lines near the section of the trajectory corresponding to the observation time $\tau$ = 0 for the parameter $\alpha = 5$ (in Fig. 6 it is the upper part of the "eight", as the y-axis of the laboratory coordinate system in the figures created in computer graphics is directed downwards). Let us compare this picture with the synchrotron radiation field of the motion of the particle along the circle contiguous to the trajectory (for the considered example the radius of such circle is 0.16656). Let us choose the speed of motion equal to the speed of motion of the charge in the field of the wave at the same moment of time: $\beta = 0.92592$

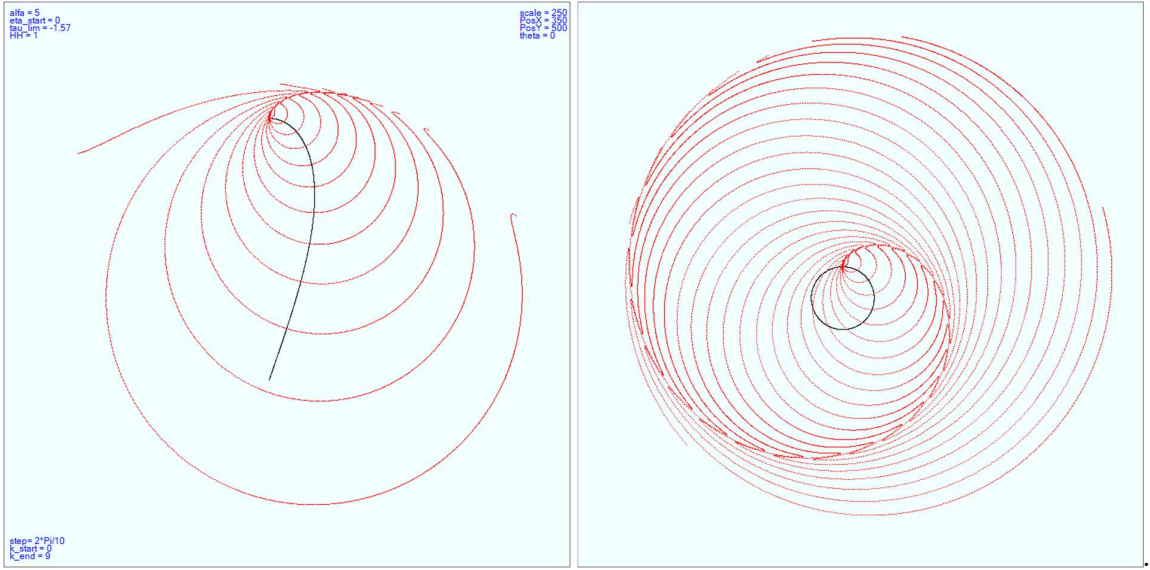

Fig. 6. Comparison of (a) the field lines for the electron motion in the wave field for $\alpha = 5$ and (b) synchrotron radiation for charge moving along the circle contiguous to the trajectory at $\tau = 0$ ($\beta = 0.92592$).

As can be seen from Fig. 6 in the region close to the charge position the structure of the lines (of field) is close to that of the synchrotron radiation.

Figs. 7-11 present electric field lines for values of parameter $\alpha = 0.1; 0.5; 1; 2; 5$. The figures are presented in the same scale and the black line represents the trajectory of the charge. The lines are drawn at time $\tau = 0$ - this is the upper point of the trajectory with maximum curvature (see Eq. (57)).

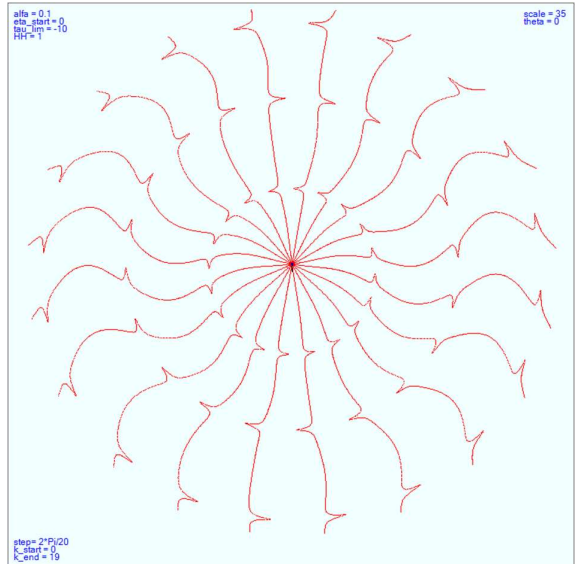

Fig. 7. Parameter $\alpha = 0.1$. On the scale of the figure the trajectory of the particle is almost invisible (small dash pointing down in the center of the picture). The field lines have in general a Coulombian character. The maximum values of velocity $\beta_{MAX} = 1.401128\text{E-}01$ and energy $\gamma_{MAX} = 1.009963\text{E+}00$ are reached at the intersection points of the "eight", while the minimum values of velocity $\beta_{MIN} = 4.975124\text{E-}03$ and energy $\gamma_{MIN} = 1.000012\text{E+}00$ are reached at the far points (along the axis $Y$) of the trajectory.

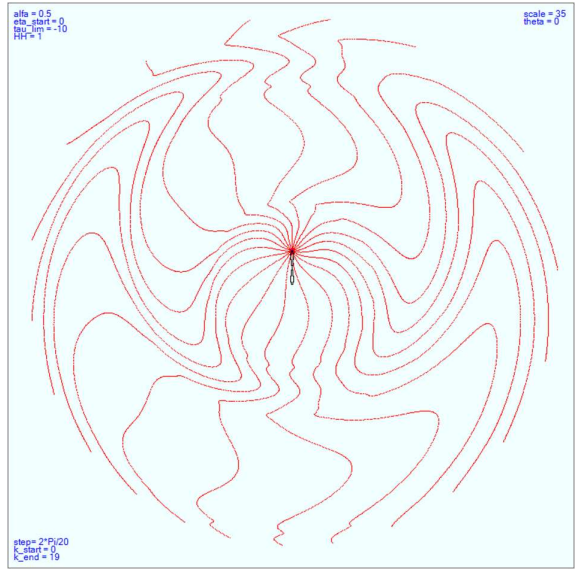

Fig. 8. Parameter $\alpha = 0.5$. The process of appearance of structures inherent to relativistic particle motions begins ($\beta_{MAX} = 5.821022\text{E}-01$; $\gamma_{MAX} = 1.229837\text{E}+00$; $\beta_{MIN} = 1.111111\text{E}-01$; $\gamma_{MIN} = 1.006231\text{E}+00$).

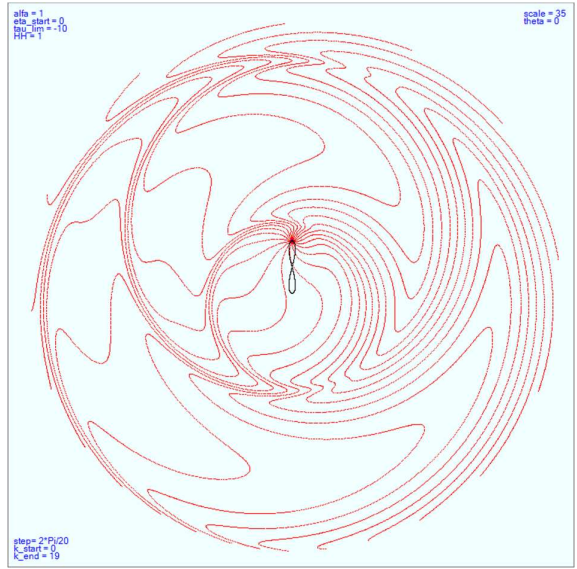

Fig. 9. Parameter $\alpha = 1$. We can see the concentration of the radiation fields in narrow spatial regions (radiation is maximal at the extremes of the trajectory, $\beta_{MAX} = 8.246211\text{E}-01$; $\gamma_{MAX} = 1.767767\text{E}+00$; $\beta_{MIN} = 3.333333\text{E}-01$; $\gamma_{MIN} = 1.060660\text{E}+00$)).

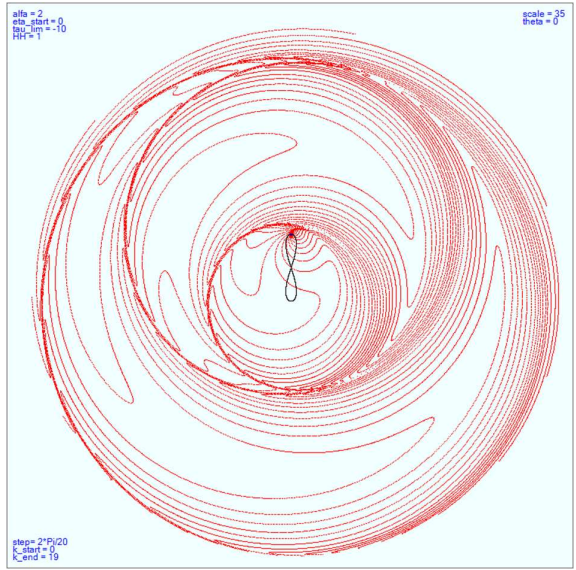

Fig. 10. Parameter $\alpha = 2$. The radiation field occupies increasingly narrower spatial regions ($\beta_{MAX}$ = 9.476071E-01; $\gamma_{MAX}$ = 3.130495E+00; $\beta_{MIN}$ = 6.666667E-01; $\gamma_{MIN}$ = 1.341641E+00).

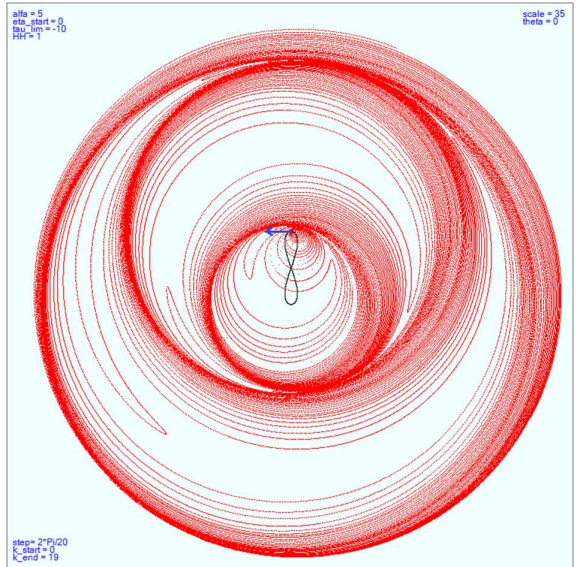

Fig. 11. Parameter $\alpha = 5$. The whole energy range of the particle on the trajectory is relativistic ($\beta_{MAX}$ = 9.911907E-01; $\gamma_{MAX}$ = 7.550471E+00; $\beta_{MIN}$ = 9.259259E-01; $\gamma_{MIN}$ =2.647568E+00). Clearly visible is the increase in the number of the lines turns around the trajectory of motion in accordance with the phase overrun analysis (see Fig. 4).

As follows from the analysis of total charge radiation intensity this value is maximal at the far points (along the axis $Y$) of the "figure eight" (in Fig. 11 the motion of the particle is shown by arrow) and is directed in the same to motion direction (from right to left). Cause of relativism of the motion radiation is concentrated in the narrow cones along speed of the particle. It is the reason of asymmetry of the whole field in horizontal direction in Fig. 11. This asymmetry is well seen in Fig. 10 and a little bit weaker in Fig. 9 too. Note that the wave propagates from left to right.

## 3.3 Field lines outside the orbital plane

The electric field lines in 3D space have a rather complex structure forming complicated surfaces. It should be noted that an attempt to represent the electric field lines of synchrotron radiation in space was made in the preprint [25]. The work was not later published and, in addition, insufficiently effective computer tools were used at that time.

The direct drawing of field lines without the analysis of the distance from sections of lines to the observer does not give a clear picture of their location in space. Thus, Fig. 12a shows a surface of lines for a charge with parameter $\alpha = 1$ and constant $H = 0.3$. The surface is woven of 90 lines corresponding to a uniform distribution of the constant $\varphi_0$ within $2\pi$. The specific caustics of the field can be seen, but the picture does not add up (the image plane ($X_{SC}$, $Y_{SC}$) is rotated with respect to the plane ($X$, $Y$) of the laboratory coordinate system by an angle $60°$, the rotation axis is $X$ axis).

In fact, to represent 3D lines/surfaces requires the use of algorithms to erase the so-called invisible sections of lines/surfaces. We have chosen a fairly simple algorithm for using a height buffer. For a drawing field (coordinates ($X_{SC}$, $Y_{SC}$)), consisting of 750x750 pixels, we generate an array of the highest heights ($Z_{SC}$) for the surface to be depicted. Only the areas corresponding to these highest heights are finally drawn.

In our case the surface is created from lines corresponding to different initial parameters $\varphi_0$, the so-called wire-form definition of the surface. In some cases the areas between such lines lend themselves to simple analysis and consequently the height buffer ($Z$-buffer) is defined analytically. In our case the lines are defined by complex parametric formulas, with a variable step algorithm for the sweep parameter (see formula (60)) being applied to avoid discontinuities in the single line representation.

Therefore, we have applied the following algorithm: $Z$-buffer is constructed with a large number of auxiliary lines, so-called construction lines, corresponding to the initial parameters $\varphi_0$ in a given range with small increments $d\varphi_0$. A set of limited number of lines is then drawn against such auxiliary lines using a different color. Several problems have come to light. First: all lines emerge from the charge and fill the pixel space quite densely at close range.

Later on, the lines begin to diverge and eventually no construction line passes through some pixels. When image lines pass through such pixels, they are deliberately drawn to form short dots or dashes. Further buffer analysis and processing appears to be required here, with such empty pixels being filled with data based on analysis of adjacent pixels.

Another problem is that many construction lines pass through some pixels and end up entering their highest coordinate value $Z$ in the corresponding buffer cell. On the other hand, if an image line is to be displayed here, but its height is slightly less than the value entered in the buffer, it becomes invisible.

We solve this problem by comparing heights with an excess of the $\delta$ value, which is chosen empirically (too high a value of this factor leads to collisions in the area of surface bends, where the real difference between the heights of the depicted and invisible parts of the surface is really small).

In order to implement this algorithm, we have developed a program in VB10. The angle of rotation of the image plane ($X_{SC}$, $Y_{SC}$) with respect to the plane ($X$, $Y$) of the laboratory coordinate system is used as the 3D parameter (rotation is performed around the axis $X$). The value of constant $H = 1$ corresponds to the image of the plane of charge motion).

The basic steps of the image are as follows: setting the parameter $\alpha$, constant $H$, constant range $\varphi_0$, as well as the angle $\theta$. Without applying the algorithm of erasing invisible areas, several lines in the area of the desired depth of immersion in the past ($\tau_{LIM}$) are constructed and the parameters of scale $scl$ and positioning of the drawing are selected. Further, the parameter of the number of construction lines is set and the buffer is filled. Usually, the number of such lines is chosen within several thousands. The area of buffer pixels on the screen is shown in yellow-green color.

Missing pixels in the main background of the drawing indicate that some pixels are not covered by design lines and it is desirable to increase their number. Next, you choose the number of lines and their color. The constructed drawing is saved in png format. The results of height analysis may be saved in a dynamic xlsx file which should be copied and renamed for further storage. The process of building one image takes quite a long time (several minutes), so dynamic rotation of these pictures as it is usually used in modern CAD programs is not yet implemented.

The results of this analysis are shown in Fig. 12b for the same values of parameters $\alpha = 1$ and constant $H = 0.3$ and 90 image lines in the range $\varphi_0$ from 0 to $2\pi$. The picture begins to become clearer, but remains rather confusing, in particular, it is difficult to imagine how the field is coupled to the trajectory of the particle.

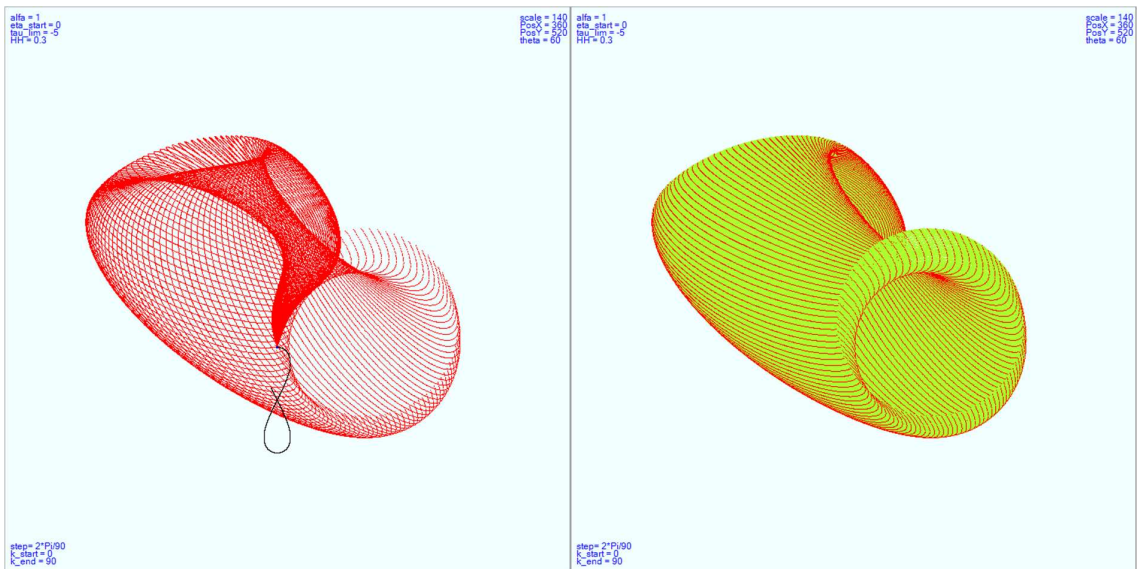

Fig. 12. The surface $H = 0.3$ ($\theta = 60°$). (a) 90 electric field lines without the procedure of erasing of invisible parts of the surface (the black color shows the trajectory of the particle). (b) The same 90 lines are depicted using the algorithm for erasing invisible parts of the surface.

In order to find out this structure we have drawn surfaces corresponding to a limited range of the parameter $\varphi_0$. Fig. 13 shows such an attempt for $H = 0.6$.

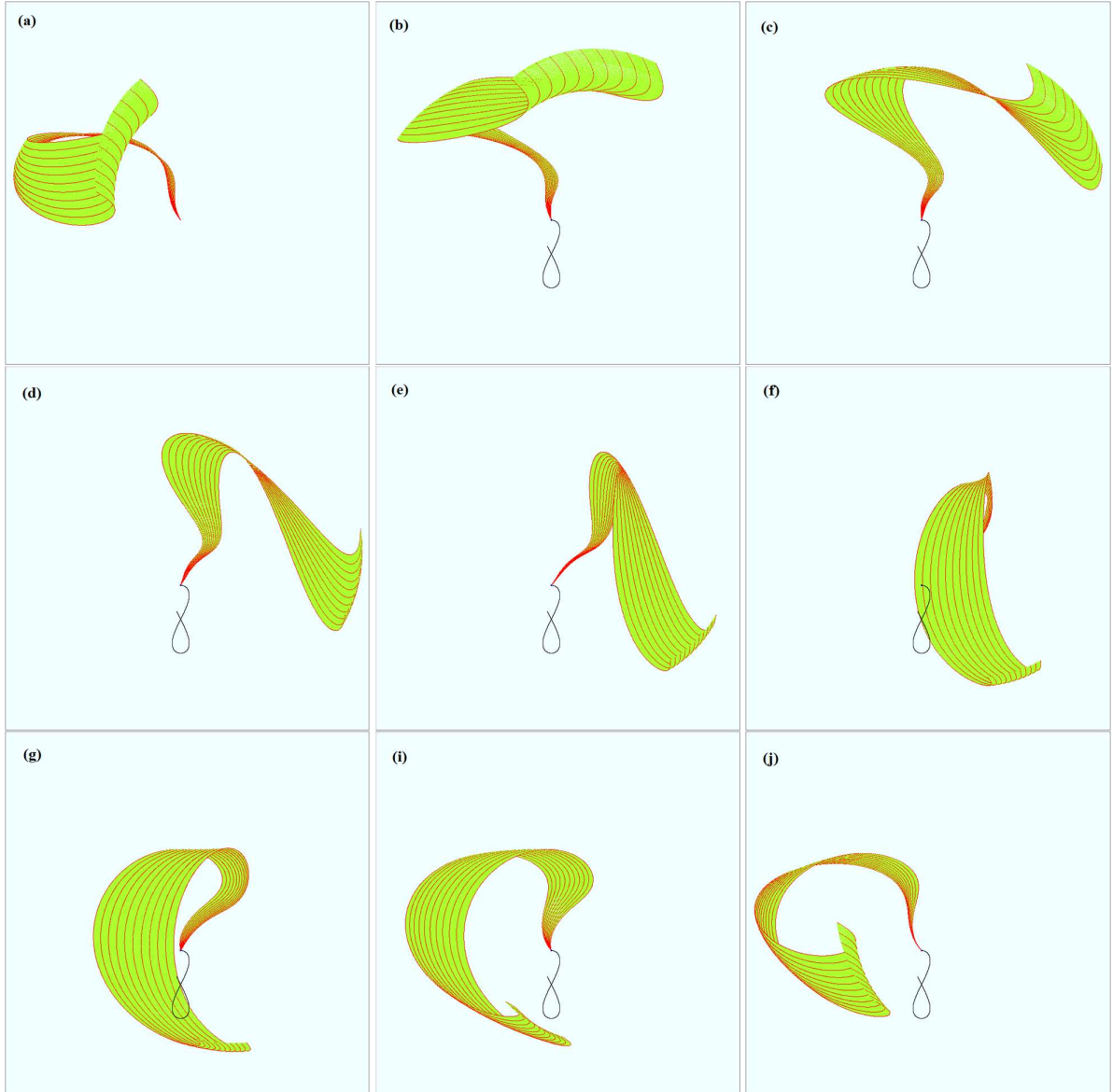

Fig. 13. The nine surface areas corresponding to the parameter $H = 0.6$ ($\theta = 60°$) are presented. The ranges of the parameter $\varphi_0$: $((0, 2\pi/9)$ (a); $(2\pi/9, 4\pi/9)$ (b); $(4\pi/9, 6\pi/9)$ (c); $(6\pi/9, 8\pi/9)$ (d); $(8\pi/9, 10\pi/9)$ (e); $(10\pi/9, 12\pi/9)$ (f); $(12\pi/9, 14\pi/9)$ (g); $(14\pi/9, 16\pi/9)$ (h); $(16\pi/9, 2\pi)$ (i)) are chosen so that the whole surface is eventually filled. The trajectory of the particle is shown in black.

And additionally: Fig. 14 shows, surfaces woven from field lines for the same parameter range $\varphi_0$: $(9\pi/5, 2\pi)$ while the parameter $H$ was chosen: 1, 0.97, 0.9, 0.8, 0.7.

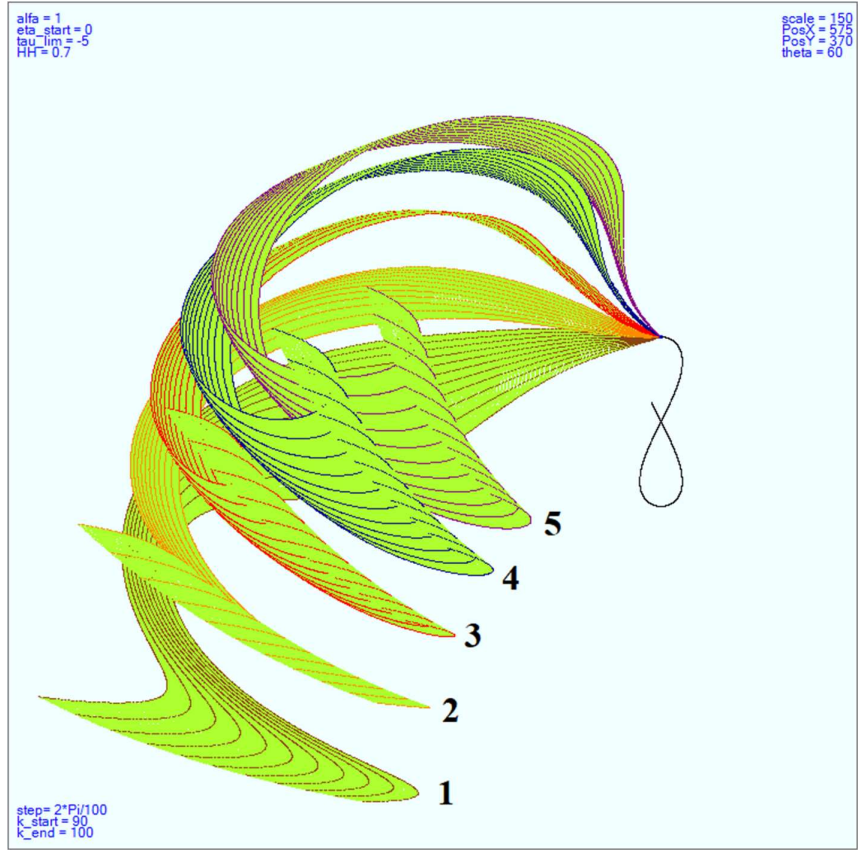

Fig. 14. Several surfaces are represented for $H = 1$ (1, orbital plane); 0.97 (2); 0.9 (3); 0.8 (4); 0.7 (5). In all cases, the construction lines were depicted for $\varphi_0$ in $(9\pi/5, 2\pi)$ range (surface marked by yellow-green) and the ten pictorial electric field lines in the same range. Surface (1) lies in the plane of the charge trajectory (shown in black).

**4. Conclusion**

This paper shows that the equations of electric field lines of an arbitrarily moving charge are reduced to linear differential equations. These equations are solved analytically in some particular cases, in particular, for any planar motion. In this case, the field analysis is performed both on the plane of motion and outside it. Such visualization of the field, especially in the case of relativistic particle motion, is useful for its representation, because it clearly indicates the fine structure of the field, including the spatial concentration of radiation. It is important that the field picture is determined at any distance from the particle and thus the process of radiation formation near the particle is well visible. Such information can be useful for the field of a bunch of particles, which consists of the sum of fields of separate particles. The nature of this superposition strongly depends on the density of the particles in the bunch and may also strongly depend on the density structure of the beam. It seems that this information should be taken into account in averaging of the bunch fields, which is often performed in beam physics and in acceleration physics, and leads to loss of rigid components of the summary field.

**5. Acknowledgments**

The work was greatly motivated by the activity and papers of B.M. Bolotovskii.
This research was supported by RA MES SCS in the frame of project 20APP-2G001.